\documentclass[a4paper,11pt]{article}
\usepackage{graphicx}
\usepackage{lineno}
\usepackage{authblk}  
\usepackage{cite}

\begin{document}
\markboth{Krystian Roslon}{Coaching station for FIT on-call shifters}


\title{Coaching station for FIT on-call shifters}

\author{Krystian Rosłon for the ALICE collaboration}
\affil{Warsaw University of Technology\\ \texttt{krystian.roslon@pw.edu.pl}}



\maketitle

\begin{abstract}
We present an efficient and cost-effective way to train operators of complex systems with limited accessibility, such as sub-detectors of big experiments at CERN Large Hadron Collider (LHC). Our coaching station was developed to train on-call shifters for the ALICE Fast Interaction Trigger (FIT). This coaching station significantly reduces the training period and enhances the trainee’s confidence in their actions. We hope this work will inspire the construction of comparable training systems for other sub-detectors.

\end{abstract}

\section{Motivation}	

Big experiments result from decades of work by thousands of scientists, engineers, technicians and students. The challenges encountered and the solutions implemented are mainly specific to these projects and, therefore, require dedicated hands-on training to gain operational proficiency. Data taking during LHC operation occurs on a 24/7 schedule. Thus, there is very little time for such training. We designed and built a coaching station to train experts and prepare a sufficient pool of well-trained on-call shifters for the ALICE-FIT detectors. The station uses all the critical components of the actual system, including a PhotoMultiplier Tube (PMT), low and high-voltage power supply, Laser System, Front-End Electronics (FEE), First Level Processor (FLP), and the Local Trigger Unit (LTU). While our coaching station was developed exclusively for FIT, it could be modified to benefit other ALICE subsystems.

\subsection{ALICE Fast Interaction Trigger }
A Large Ion Collider Experiment (ALICE) \cite{Aamodt:2008zz} is one of the four major CERN Large Hadron Collider experiments. ALICE studies the properties of quark-gluon plasma (QGP), a state of matter where quarks and gluons are deconfined. The ALICE physics program includes proton, lead, and oxygen collisions. ALICE can track and identify particles down to low transverse momenta, allowing detailed studies of the essential properties of the QGP. 

\begin{figure}[h]
\begin{center}
\includegraphics[width=9.5 cm]{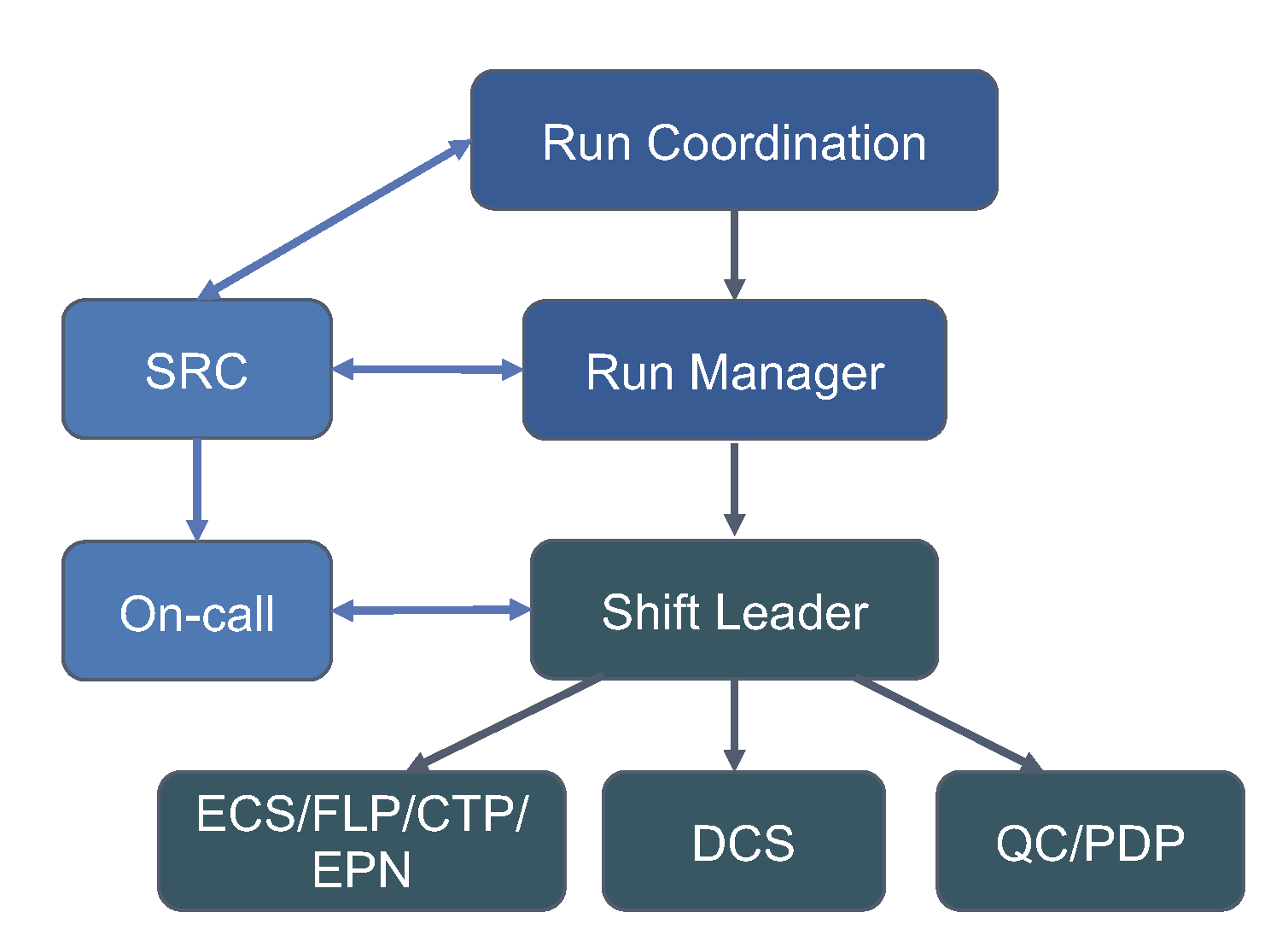}
\caption{ALICE Run Coordination structure\cite{Alice-operation}. All used abbreviations are listed at the end of this document.\label{shift}}\end{center}
\end{figure}

FIT \cite{Trzaska:2020zzl} detector has been added to the ALICE setup during the Long Shutdown 2 (2019 -- 2022) and will operate throughout Run 3 (2022 -- 2026) and 4 (2030 -- 2033). FIT is the first detector turned on, and the last turned off, during data-taking. It provides online luminosity, precise collision time, initial vertex position, fast triggers, forward multiplicity, event plane and centrality. Since FIT is essential to the operation of ALICE, it must be constantly monitored and promptly fixed/corrected in case of emerging issues. That responsibility is with the ALICE Run Coordination (RC) schematically depicted in Fig. \ref{shift}. In case of problems, the Shift Leader contacts relevant on-call experts. Since data-taking covers most of the year with 24/7 operations, we need a sufficiently large pool of well-trained FIT experts to meet that challenge.

\section{Development of the FIT Training Station}
\begin{figure}[h]
\begin{center}
\includegraphics[width=12.5 cm]{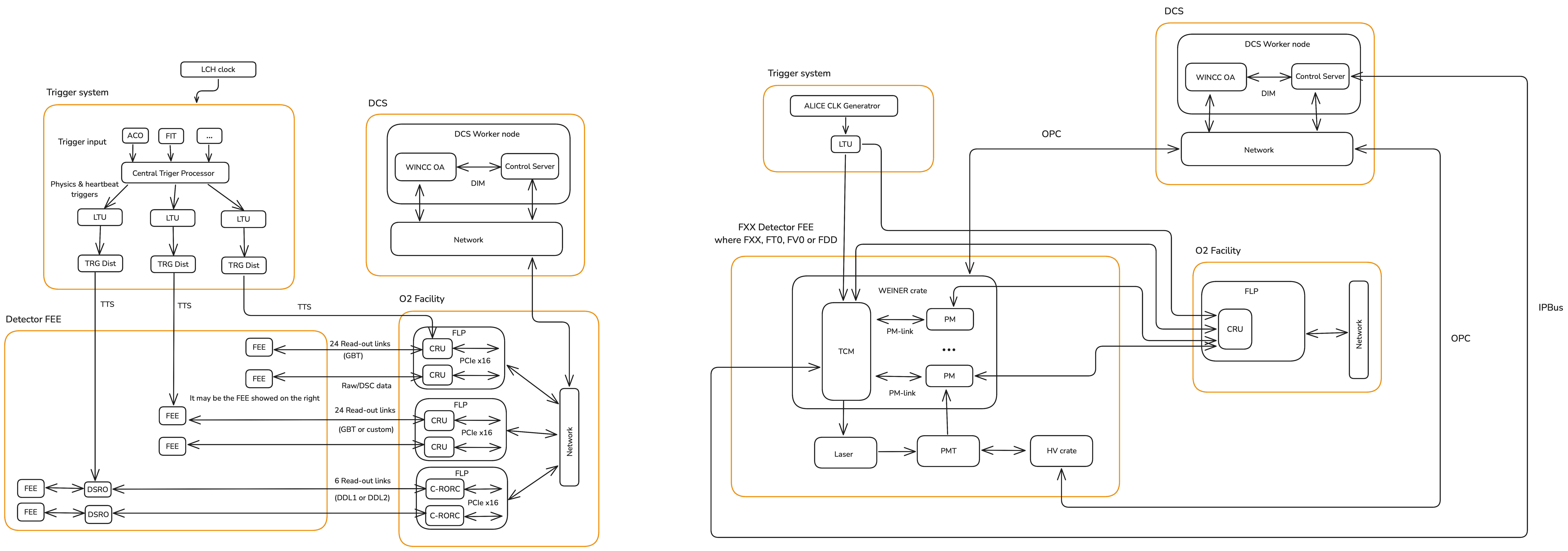}
\caption{Left: Detector read-out and interfaces of the O2 system with the trigger, detector electronics and DCS. Right: Schema of the FIT laboratory setup.\label{readout}}\end{center}
\end{figure}

Training new people to operate and maintain FIT requires hands-on work experience with all the critical components of the detector and its control. Therefore, such practice must be scheduled outside the data-taking periods. However, these beam-off periods are infrequent and often brief due to the LHC’s tightly coordinated operating schedule \cite{LHC_Schedule2024}. The only exception is the Year End Technical Stop (YETS). Since YETS overlaps with the Christmas Holidays and the annual CERN closure it is unsuitable for training. These were the main reasons to develop a setup that can be used anytime throughout the year. Figure \ref{readout} illustrates the transition from the operational FIT detector to the lab-based training station. This simplification retained FIT’s essential components while providing a robust, scalable environment to train new on-call experts during the ALICE data-taking periods.

\subsection{Hardware}
The hardware configuration of the FIT Training Station replicates the core architecture of the FIT detector as implemented in the ALICE experiment at CERN.

At the system's core is the PMT, a highly sensitive device that converts light signals into electrical signals for further processing. The PMT is powered by a CAEN High Voltage (HV) Crate, providing precise and stable voltage needed for signal amplification. The PMT is housed in a custom-designed, light-tight box connected to a laser system. The laser system is integrated with the FIT Front-End Electronics. FEE includes one Trigger and Clock Module (TCM) and two Processing Modules (PMs). TCM receives the status data from the control system and triggers the laser. The data is read and digitised by the two PM units.
All system components are centrally controlled via a WinCC and Control Server software interface installed on a dedicated computer. This software allows users to configure, monitor, and control all aspects of the hardware, ensuring seamless operation of the training station.
The TCM communicates with the FLP node, which contains the Common Readout Unit (CRU). The CRU is responsible for data acquisition and processing via the GigaBit Transceiver (GBT) protocol.
In the case of the training station, the LTU is connected to the FLP to provide clock signals, thereby simulating the internal ALICE clock. This is crucial for accurately reproducing the actual operating environment of the detector, such as switching the clock from the internal (FIT) to the external (ALICE - LTU) source.

\begin{figure}[h]
\begin{center}
\includegraphics[width=12.5 cm]{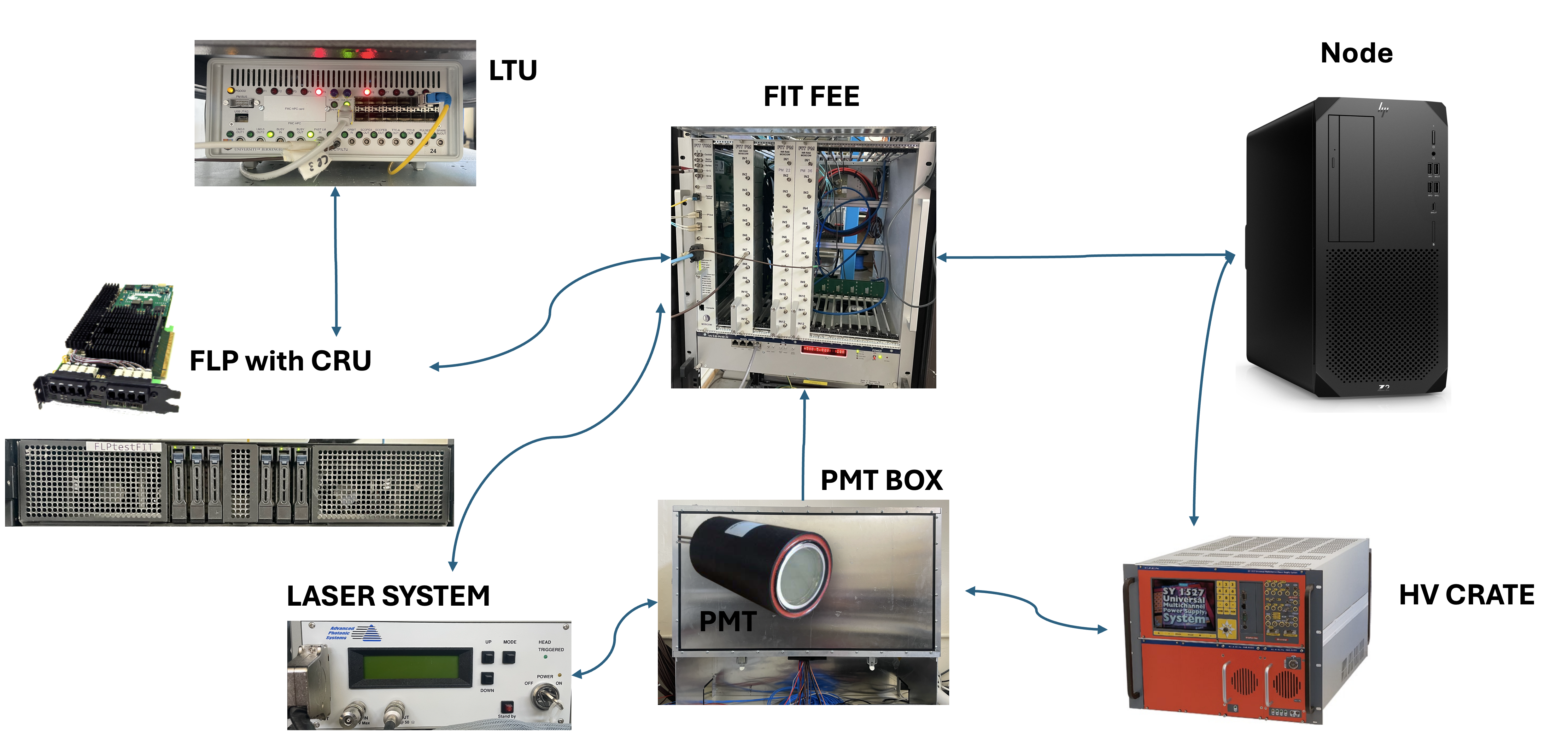}
\caption{Components of the ALICE FIT coaching station.\label{lab}}\end{center}
\end{figure}

Since the training station contains the key FIT components and processes, trainees can familiarise themselves with the detector’s operations in a controlled environment. In particular, they can interact with the system just as they would during a live data collection run. (See Fig. \ref{lab})

\subsection{Software}

The software component of the FIT Training Station is centred on the Control Server, a QT-based application running on a dedicated "Node" computer. This system communicates with the FIT FEE via the IPbus protocol and provides extensive functionality for monitoring, managing, and controlling the detector subsystems. Additionally, all visualisation and integration with the central ALICE-DCS system is facilitated using WinCC O.A. 3.19 \cite{Rodriguez}, which is responsible for the overall operation of the system and launched on the same “Node” computer.

Core Functions of the Control Server:
\begin{enumerate}
  \item Monitoring and Diagnostics: Periodic polling of all connected FEE modules to monitor their state and detect errors, with detailed logging for diagnostics.
  \item Error Handling: Verification and blocking of erroneous TCM and PMs register readings, with real-time error notifications for individual modules and the overall system.
  \item Configuration Management: Display and management of TCM and PM settings, including trigger outputs, laser systems, attenuators, and detector channels.
  \item DIM Integration:
  \begin{itemize}
      \item Publication of DIM services for counter readings, trigger rates, and parameter settings.
  \end{itemize}
  \item Logging and Diagnostics:
  \begin{itemize}
      \item Comprehensive event logging, including critical errors and diagnostic reports such as “BC sync lost in run” and “raw data FIFO overload.”
      \item Separation of critical error logs into ControlServer.errorlog for efficient troubleshooting.
  \end{itemize}
  \item GUI Enhancements:
  \begin{itemize}
      \item Intuitive display of system states and parameters, with input validation and tooltips for user-friendly operation.
      \item Real-time visualisation of average signal registration times for each detector side, with quick correction options for temporal delays.
  \end{itemize}
\end{enumerate}

Key Tasks Performed by WinCC O.A. 3.19:
\begin{enumerate}
    \item Control of Hardware Components:
    \begin{enumerate}
        \item Wiener crates (via OPC server).
        \item CAEN crates (via OPC server).
    \end{enumerate}
    \item Communication with Control Server:
    \begin{enumerate}
        \item Exchange of DIM services and commands for interaction with the TCM.
    \end{enumerate}
    \item Finite State Machine (FSM) Management:
    \begin{enumerate}
        \item Creation of an FSM to define the operational states of the detector.
        \item Development of a safety matrix mapping FSM states to ALICE BEAM Safe or Magnet Safe states.
    \end{enumerate}
    \item Visualisation:
    \begin{enumerate}
        \item Display of configuration parameters such as currents, voltages, resistances, and data sent to the LHC interface (e.g., trigger information).
    \end{enumerate}
    \item Data Archiving:
    \begin{enumerate}
        \item Comprehensive logging and archival of all relevant operational and configuration data.
    \end{enumerate}
    \item Alert Management:
    \begin{enumerate}
        \item Definition of alert thresholds, including Warning and Error levels, and provision of real-time alerts.
    \end{enumerate}
\end{enumerate}

\begin{figure}[h]
\begin{center}
\includegraphics[width=9.5 cm]{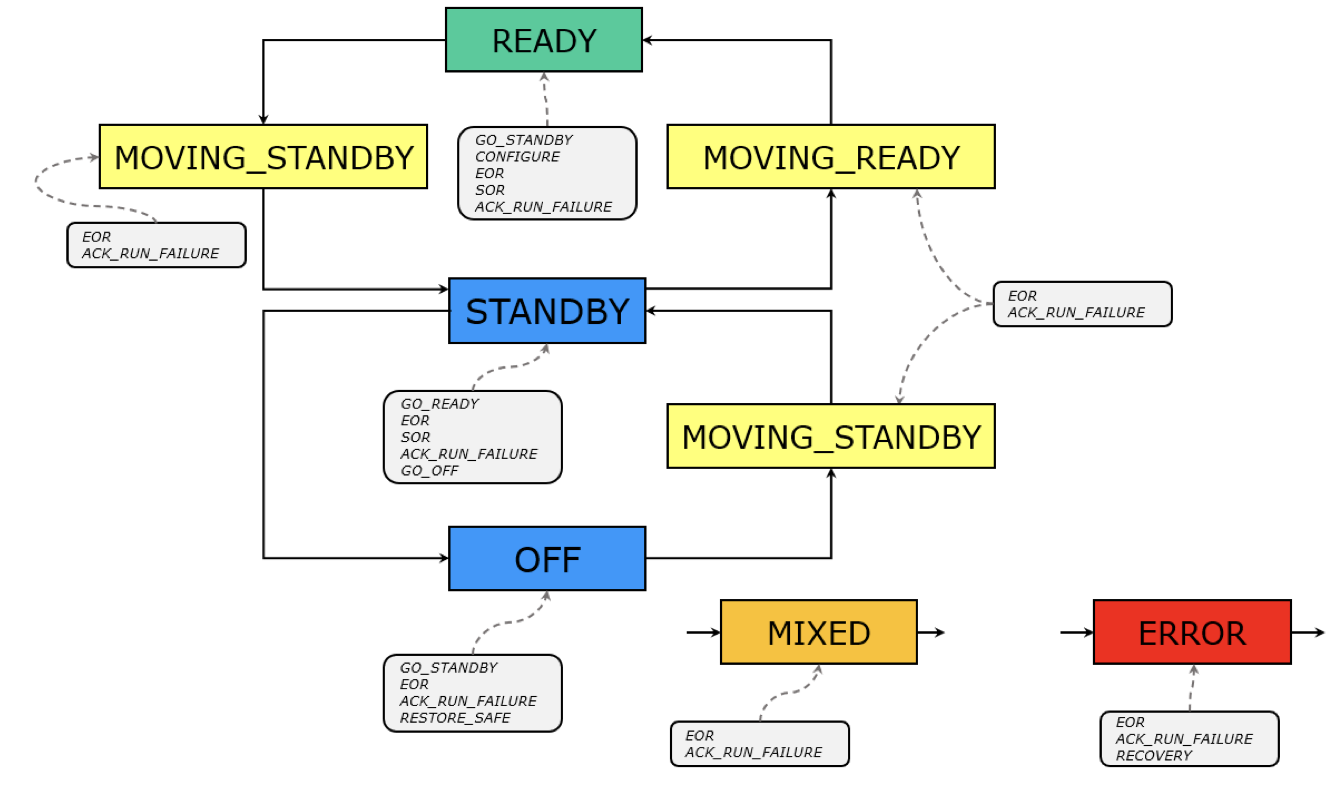}
\caption{ALICE FIT Finite State Machine layout, details available\cite{Rodriguez}.\label{FSM}}\end{center}
\end{figure}

The FIT detector system is unified, consisting of three sub-detectors that utilize very similar electronics, read-out, and a comparable control and monitoring system.\cite{Slupecki:2022fch} The Win-CC O.A. \cite{WinCCOA} FSM is also standardized across all sub-detectors and coaching station \cite{MejiaCamacho:2023zam} (see Fig. \ref{FSM}). This cohesive design of the FIT system is reflected in the training station, which incorporates unified interfaces and controls that emulate the operational environment closely.

\begin{figure}[h]
\begin{center}
\includegraphics[width=8.5 cm]{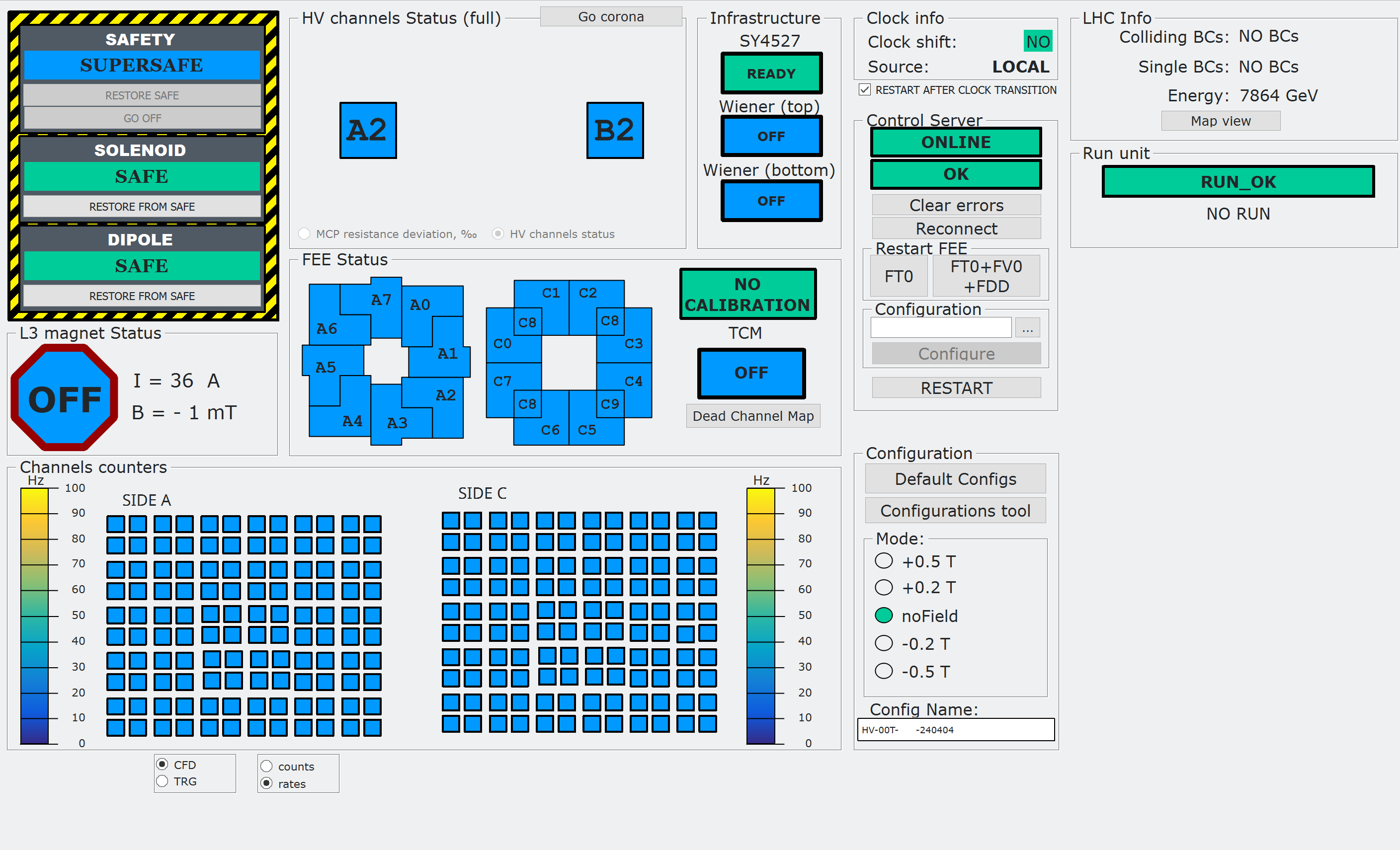}
\caption{Supervisory Control And Data Acquisition (SCADA) of ALICE FIT laboratory setup\label{WinCC}.}\end{center}
\end{figure}

The training station is in the FIT laboratory, where spares of the system’s critical components are kept as well. They include electronics, data acquisition systems, the laser calibration system, and backups of essential software. So, the trainees have direct access not only to the key hardware elements but also to the software. A comprehensive user interface was developed to mimic the control panels used by on-call experts. (See Fig. \ref{WinCC}) This allows trainees to navigate the system’s	 controls, monitor real-time data streams, and execute typical actions that would be required during live operations.

\section{Benefits of the Training Station}

Unlike traditional training methods that could interfere with data collection during LHC run periods, the station operates independently from the main detector. This separation allows trainees to engage in hands-on practice without risking disruptions to ALICE’s data collection or overall performance. The ability to train without interrupting operations is a major benefit, particularly in the high-stakes environment of the LHC, where maintaining continuous data collection is of paramount importance.

Another advantage is the substantially reduced time required to train new FIT experts. Previously, the limited opportunities for hands-on experience—restricted to beam-off periods or technical intervention windows—stretched the training timeline and delayed the integration of new experts into the on-call roster. With the training station in place, trainees can now undergo comprehensive practical training without waiting for these scarce windows.
The third benefit is improving trainees’ confidence and readiness to handle real-time operations. Trainees gain hands-on experience in various operational scenarios by practising in a realistic environment, including potential errors and system failures. This experience enhances their ability to troubleshoot and resolve issues independently during live operations, reducing reliance on seasoned experts. With repeated exposure to the system’s functionality and simulated problem-solving, new FIT on-call experts are better prepared to manage the complexities of detector problems during actual operation. This, in turn, minimises the response time when resolving detector issues during the run and thus increases ALICE data-taking efficiency.

\section{Conclusion and Further Works}

The development and implementation of the FIT training station has a noticeable impact on FIT on-call training and, therefore, on ALICE operational efficiency during LHC Run 3. The station addresses a critical need for hands-on training in an environment where opportunities for such experience are limited due to the high demand for continuous data collection. By enabling new on-call experts to practice and familiarise themselves with the FIT detector without interfering with live operations, the training station significantly shortens the training period and enhance the abilities and confidence of the trainees.
The introduction of this training station contribute to the smooth operation of ALICE during Run 3. It alleviates the workload of detector experts by allowing the delegation of 24/7 on-call shifts to a broader group of trained individuals. Trainees can now gain essential experience in a controlled and realistic environment, equipping them to manage real-time operations with minimal reliance on senior experts.
Looking beyond ALICE and the FIT detector, this training station model presents a compelling case for adoption by other sub-detectors at the LHC and other large-scale experiments. As high-energy physics experiments grow in complexity, the need for well-prepared on-call experts become increasingly important. Collaboration with other experiments could also lead to developing a standardised training framework that can be adapted to the unique needs of different detectors and experimental setups.

\section{Acknowledgements}
This work was supported by the Polish Ministry of Science and Higher Education under agreements no. 5452/CERN/2023/0, 2023/WK/07 and "The Excellence Initiative - Research University" program.
\section{Abbreviations}
The following abbreviations are used in this manuscript:\\

\noindent 
\begin{tabular}{@{}ll}
LHC & Large Hadron Collider \\
ALICE & A Large Ion Collider Experiment \\
FIT &  Fast Interaction Trigger \\
TOF & Time of Flight \\
SRC & System Run Coordinator \\
ECS & Experiment Control System \\
FLP & First Level Processor \\
CTP & Central Trigger Processor \\
EPN & Event Processing Nodes \\
DCS & Detector Control System \\
QC & Quality Control \\
PDP &  Physics Data Processing \\
CRU & Common Readout Unit \\
LTU & Local Trigger Unit \\
ALICE CLK & ALICE CLOCK \\
TCM & Trigger and Clock Module \\
PM & Processing Module \\
PMT & Photomultiplier tube \\
HV & High Voltage \\
CS & Control Server \\
WinCC OA & WinCC Open Architecture\\
DIM & Distributed Information Management System \\
EOR & End of Run\\
SOR & Start of Run \\
SCADA & Supervisory Control And Data Acquisition \\

\end{tabular}

\bibliographystyle{ws-ijmpa}
\bibliography{bibliography}
\end{document}